\documentclass[aip,jcp,reprint,amsmath,amssymb]{revtex4-2}
\usepackage{color}
\usepackage{graphicx}
\usepackage{comment}
\usepackage{amssymb,amsmath}

\newcommand\rev[1]{{\color{black} #1}}
\newcommand\pb[1]{{\color{blue} #1}}

\newcommand{\p}[1]{\left( #1 \right)}
\newcommand{\bk}[1]{\left[ #1 \right]}
\newcommand{\w}[1]{\mathbf{#1}}
\newcommand{\parfr}[2]{\frac{\partial #1}{\partial #2}}

\newcommand{\CC}{\mathbb{C}}
\newcommand{\EE}{\mathbb{E}}
\newcommand{\h}{\hspace{0.3em}}

\newcommand{\R}{{R_{\rm e}}}

\begin{document}


\title{Models for polymer dynamics from dimensionality reduction techniques} 



\author{Phillip Bement}  
\affiliation{Dept.~of Physics and Astronomy and Stewart Blusson Quantum Matter Institute, University of British Columbia, Vancouver, British Columbia V6T 1Z1, Canada}

\author{J\"org Rottler}
\email{jrottler@physics.ubc.ca}
\affiliation{Dept.~of Physics and Astronomy and Stewart Blusson Quantum Matter Institute, University of British Columbia, Vancouver, British Columbia V6T 1Z1, Canada}


\date{\today}

\begin{abstract}
Polymer dynamics is analyzed through the lens of linear dimensionality reduction methods, in particular principal (PCA) and time-lagged independent component analysis (tICA). For a polymer undergoing ideal Rouse dynamics, the slow modes identified by these transformations coincide with the conventional Rouse modes. When applied to the Fourier modes of the segment density, we show that tICA generates dynamics equivalent to dynamic self-consistent field theory (D-SCFT) with a wavevector-dependent Onsager coefficient and a free energy functional subject to the random phase approximation (RPA). We then introduce a hidden variable method and a time-local approach to include temporal memory in the tICA-generated dynamics, and generalize it to construct continuum models for the nonequilibrium case of spinodal decomposition of a symmetric diblock copolymer melt.
\end{abstract}


\maketitle 

\section{Introduction}
Macromolecular systems feature a rich dynamics due to a hierarchy of inherent timescales. At the level of monomer or segmental positions, an analysis in terms of Rouse modes \cite{Rouse53} (equivalent to a discrete cosine transform of the monomer positions) is typically employed to reveal the relaxation dynamics of the polymer across scales ranging from segmental spacing to the size of the polymer coil. A key result for unentangled polymer liquids is that segments exhibit normal diffusion on timescales larger than the Rouse time $\tau_R$, while \rev{subdiffusive} behavior is found on shorter timescales, reflecting intrachain dynamics. For larger molecular weights, entanglement effects intervene, and normal diffusion is only observed after chains have escaped from locally confining tubes, a notion made precise by the tube model of polymer physics \cite{degennes79book,doi1988theory}.

An alternative approach to studying in particular the dynamics of a collective order parameter is in terms of the segmental density field, or rather its Fourier components whose correlation functions are revealed by scattering experiments. Since the density is a coarse-grained function of monomer positions, it however exhibits memory effects that become particularly pronounced on short timescales.

These reciprocal space methods, where the basis functions are cosine functions or plane waves, resp., are not the only possible transformation that might reveal important aspects of the dynamics. From a more data centric perspective, principal components analysis (PCA) is frequently employed as transform into a basis with principal components that maximize sample covariances. In polymer science, PCA approaches have been invoked recently in the context of the dynamics of bottlebrushes in dilute solution \cite{mukkamala_2024_simulation} and for estimating the glass transition temperature in coarse-grained \cite{banerjee2023data} and all-atom simulations \cite{banerjee2023determining}. PCA also has a long history in the characterization of conformational shape fluctuations of proteins \cite{hayward1995collective} and DNA \cite{cohen2007principal}.

A generalization of this concept to dynamical systems is time-lagged independent component analysis (tICA), which seeks a basis of maximal autocorrelation between dynamical variables \cite{schwantes2015modeling}. In the field of biomolecular simulations, tICA has found numerous applications for identifying dominant (i.e.~slow) dynamical modes that correspond to significant conformational changes in the underlying free energy landscape \cite{m2017tica,mcgibbon2017identification,schultze2021time}.  Moreover, the tICA operator can be taken as a (finite-dimensional) approximation of the so-called Koopman operator that advances expectation values of observables in time in a linear, Markovian picture \cite{klus2018data}. 

The purpose of this contribution is to explore how PCA and tICA can aid the analysis of classical polymer dynamics. We first study the dynamics of a single Rouse polymer at equilibrium, and show that in this context the PCA and tICA transformations of monomer positions are exactly equivalent to the standard Rouse mode analysis. We then apply tICA to the Fourier modes of the segmental density field, and show that the resultant Koopman operator generates dynamics equivalent to what the literature refers to as dynamical self-consistent field theory (D-SCFT)\cite{fraaije1997dynamic,qi2017dynamic,muller2018continuum} for an ideal gas of noninteracting Rouse chains. Since tICA/D-SCFT can only describe Markovian dynamics, we develop an extension of tICA by introducing hidden variables that is capable of describing the nonexponential relaxation dynamics and memory effects in the short-time regime. The method is general and can be applied to any dynamical system with memory. Finally, we present a generalized tICA analysis in a nonequilibrium, nonlinear setting and apply it to spinodal decomposition in a diblock copolymer melt.  

\section{PCA and tICA}
We consider a time series or trajectory (column) vector $\mathbf{X}(t)$ obtained for instance from a molecular dynamics simulation. First, we center this trajectory by subtracting off the mean,
\begin{equation}
    \mathbf{\bar{X}}(t)=\mathbf{X}(t)-\langle \mathbf{X}(t) \rangle
\end{equation}
Next, we define the equal-time covariance matrix
\begin{eqnarray}
    C_{00}&=&\langle  \mathbf{\bar{X}}(t) \mathbf{\bar{X}}(t)^T \rangle_t = \EE[\mathbf{\bar{X}}(t)\mathbf{\bar{X}}(t)^T].
\end{eqnarray}
Principal component analysis (PCA) computes the spectrum of this (real symmetric) matrix, 
\begin{equation}
    C_{00}\mathbf{v}_i=\lambda_i\mathbf{v}_i,
\end{equation}
where $\mathbf{v}_i$ denote the eigenvectors of $C_{00}$. A standard result is that for positive semidefinite matrices, the Rayleigh quotient 
$$R(\mathbf{v})=\frac{\mathbf{v}^T C_{00} \mathbf{v}}{\mathbf{v}^T\mathbf{v}}$$
is maximized by the eigenvector with the largest eigenvalue. As a result, the first principal component of the transformed data has maximal variance, and the eigenvalue can be obtained from a variational principle.

PCA looks only at the instantaneous covariances and does not reveal any dynamical information. For dynamical systems, a more interesting quantity is the time lagged covariance  
\begin{eqnarray}
    C_{0\tau}&=&\langle  \mathbf{\bar{X}}(t) \mathbf{\bar{X}}(t+\tau)^T \rangle_t
\end{eqnarray}
This, however, should be computed for standardized or whitened coordinates, whose equal time covariance matrix is diagonal and the components are thus decorrelated and mutually orthonormal. By definition, this transformation must be
\begin{equation}
    \mathbf{\bar{X}}^{'}(t)=C_{00}^{-1/2}\mathbf{\bar{X}}(t)
\end{equation}
so that 
\begin{equation}
    \langle  \mathbf{\bar{X}}^{'}(t) \mathbf{\bar{X}}^{'}(t+\tau)^T \rangle_t=C_{00}^{-1/2}C_{0\tau} C_{00}^{-1/2}
\end{equation}

We now seek a transformation that maximizes the autocorrelation instead of the variance. This is achieved by maximizing the generalized Rayleigh quotient
$$R(\mathbf{w})=\frac{\mathbf{w}^T C_{0\tau} \mathbf{w}}{\mathbf{w}^TC_{00}\mathbf{w}}.$$ 
The desired transformation is found by solving the generalized eigenvalue problem
\begin{equation}
    C_{0\tau}\mathbf{w}_i=\lambda_iC_{00}\mathbf{w}_i.
\end{equation}

Equivalently, one diagonalizes the matrix $K(\tau)=C_{00}^{-1}C_{0\tau}$. By convention, the transpose $K^T(\tau)=C_{0\tau}^T C_{00}^{-1}=C_{\tau 0} C_{00}^{-1}$ is the (finite-dimensional) time-lagged Independent Component Analysis (tICA) estimate of the (infinite-dimensional) Koopman operator $\mathcal{K}$, which propagates expectation values of observables $f$\cite{klus2018data,schutt2020machine}, i.e
\begin{equation}
    \mathcal{K}_\tau f(x)=\EE[f({\bf X}(t+\tau)|{\bf X}(t)=x]
\end{equation}

\section{Equilibrium Dynamics of Rouse polymer segments}

For a Rouse model polymer of $N$ beads, PCA and tICA turn out to be entirely equivalent to a conventional Rouse mode analysis. In order to see this, it suffices to consider a chain in one dimension. The Rouse modes $u_0, u_1, \dots u_{N-1}$, given by taking a discrete cosine transform (DCT) of the bead coordinates, are:
\begin{equation}
    u_{n} =
\sqrt{\frac{2 - \delta_{0,n}}{N}} \sum_{m=0}^{N-1}
\cos\p{\frac{\pi (m+\frac{1}{2}) n}{N}} x_m
\end{equation}
where $x_m$ gives the position of the $m$th bead. With overdamped dynamics, the polymer's equation of motion is

\begin{equation}
\parfr{\w{X}}{t} = -K_S\w{X} + \parfr{\w{W}}{t}
\end{equation}
where $\w{X}$ is the vector of the coordinates of all the beads, $\w{W}$ is a Wiener process, and $K_S$ is the spring matrix for the chain. $K_S$ can be diagonalized as: $K_S = U^T \Lambda U$ where $U$ is the DCT matrix (i.e. $u_n = (U\w{X})_n$), and the eigenvalues are
\begin{equation}
    \Lambda_{nn}= 4\sin^2\p{\frac{n\pi}{2N}}
\end{equation}
(In 3 dimensions a 3-fold degeneracy is introduced, but the discussion is otherwise unchanged.) The equilibrium distribution $q(\w{X})$ should be a fixed point of the system's Fokker-Planck equation:
\begin{equation}
    \parfr{q(\w{X})}{t} = \nabla\cdot\p{K_S\w{X}\h q(\w{X})} + \frac{1}{2}\nabla^2 q(\w{X}) = 0
\end{equation}
Ignoring normalization, $q(\w{X}) \sim \exp(-\w{X}^T K_S \w{X})$ solves this, and it is a Gaussian distribution which has covariance
\begin{equation}
C_{00} = \frac{1}{2}K_S^{-1} = \frac{1}{2} U^T \Lambda^{-1} U
\end{equation}
From this we can see that the principal components are the Rouse modes, and these modes have variances $\sigma_n = 1/4\sin^2\p{\frac{n\pi}{2N}}$. Of particular interest is the 0th mode, which has infinite variance at equilibrium, corresponding to center-of-mass diffusion. 

To perform tICA we would next compute the time lagged covariance matrix $C_{0\tau}$.

$$
    \parfr{\h\EE[\w{X}_\tau | \w{X}_0]}{\tau} = -U^T\Lambda U \EE[\w{X}_\tau | \w{X}_0]
$$$$
    \EE[\w{X}_\tau | \w{X}_0] = \exp\p{-\tau \h U^T\Lambda U} \h \w{X}_0
$$
\begin{eqnarray}
    C_{0\tau} = \EE[\w{X}_0 \w{X}_\tau^T] = \EE\bk{\w{X}_0 \w{X}_0^T} \h \exp\p{-\tau \h U^T\Lambda U} \\
    = C_{00} \exp\p{-\tau \h U^T\Lambda U} = \frac{1}{2} U^T \Lambda^{-1} \exp\p{-\tau\Lambda} U
\end{eqnarray}

So the finite dimensional operator $K$ found by tICA is:
\begin{equation}
    K = C_{00}^{-1}C_{0\tau} = U^T \exp\p{-\tau\Lambda}U.
\end{equation}

When this operator is applied to the particle positions ${\w X}$, it first transforms them in into Rouse modes, then propagates each of them with characteristic time $\tau_n=1/\Lambda_{nn}$, and then transforms the Rouse modes back to physical coordinates.

Based on this analysis, the most significant (largest variance) modes identified by PCA also happen to be exactly those modes with the slowest dynamics. It is instructive to note which properties of the Rouse model are relevant to this result:
\begin{enumerate}
    \item {Linear dynamics described by a positive-semidefinite spring matrix $K_S$ (Any such matrix can be diagonalized as $K_S = U^T \Lambda U$).}
    \item {All beads have the same mass and the same friction coefficient. }
\end{enumerate}

It is not required for all springs in the chain to have the same stiffness, nor do the beads need to be connected in a linear chain as opposed to, say, a ring or star configuration. PCA and tICA will produce the same eigenmodes for any such system.

\section{Equilibrium Dynamics of the segment density}
Let us now consider the segment density 
\begin{equation}
    \rho({\bf r},t)=\sum_i^N \delta({\bf r}-{\bf r}_i(t))
\end{equation}
where ${\bf r}_i(t)$ denotes the position of the $i$th segment.
We seek to build a dynamical model with the tICA transformation. Since projection of particle positions onto the density is a coarse-graining operation, we expect its dynamics to include memory. In order to treat periodic boundary conditions, it is convenient to perform tICA not directly on the density itself, but on its Fourier components
\begin{equation}
    \rho_{\bf k}(t)=\frac{1}{V}\int d^3r e^{i{\bf k r}}\rho({\bf r},t)=\frac{1}{V}\sum_ie^{i{\bf k r}_i(t)}
\end{equation}
tICA requires us to study the spectrum of the operator $K=C_{00}^{-1}C_{0\tau}$,
where 
\begin{eqnarray}
C_{00}&=&\langle \rho_{\bf k}(t) \rho_{\bf -k'}(t)^T \rangle_t\\
C_{0\tau}&=&\langle \rho_{\bf k}(t) \rho_{\bf -k'}(t+\tau)^T \rangle_t 
\end{eqnarray}
are the equal time and time lagged covariance matrices as discussed above. For an ideal gas of Rouse chains, $C_{00}$ only contains nonzero entries for ${\bf k}=-{\bf k}'$ due to spatial averaging, while the dynamics does not mix the Fourier modes in $C_{0\tau}$. Therefore, the covariance matrices are already diagonal and correspond to the static and dynamic single chain structure factors, $S_{\bf k}(\tau)=\frac{V^2}{N}\langle \rho_{\bf k}(t)\rho_{\bf -k}(t+\tau) \rangle_t$. The diagonal entries or eigenvalues of the tICA operator $K$ are just $S_{\bf k}(\tau)/S_{\bf k}(0)$. For the Rouse chain with Gaussian chain conformations, an end-to-end distance $R_e$ and a centre-of-mass diffusion coefficient of the chain $D=R_e^2/(3\pi^2\tau_R)$, we have
\begin{equation}
    S_{\bf k}(t)/S_{\bf k}(0)=g_{\bf k}(t)/g_{\bf k}(0)
    \label{dynsf-eq}
\end{equation}
with 
\begin{eqnarray}
    g_k(t)&=&\int_0^1 ds \int_0^1 ds^\prime  \exp[-({\bf k}R_e)^2|s-s^\prime|/6-{\bf k}^2Dt \nonumber \\ &-&\frac{2({\bf k}R_e)^2}{3\pi^2}\sum_{p=1}(1-e^{-p^2t/\tau_R})/p^2\cos(\pi ps)\cos(\pi p s^\prime)] \nonumber
\end{eqnarray}
and $g_{\bf k}(0)=2(e^{-x}+x-1)/x^2$ \rev{with} $x=({\bf k}R_e)^2/6$ the Debye function \cite{Rouse53,wang2019collective}.

\subsection{Markovian dynamics for $t>\tau_R$}
For times larger than the Rouse time $\tau_R$, the decay of the dynamic structure factor of a Rouse chain is purely exponential, $S_{\bf k}(\tau)\simeq S_{\bf k}(0)\exp[-Dk^2\tau]$. As a result, $K_{kk}=\exp[-Dk^2\tau]$. The time evolution of the segment density can then be obtained by applying the tICA operator,
\begin{equation}
    \rho_{\bf k}(t+\tau)=K^T_{kk}(\tau)\rho_{\bf k}(t)=\exp[-Dk^2\tau]\rho_{\bf k}(t)
    \label{eq-ticamodel}
\end{equation}
The tICA dynamics therefore predicts an exponential decay of density fluctuations in time, as can be expected for a time-reversible Markov process. Interestingly, the same dynamics is also predicted by dynamic polymer self-consistent field theory (D-SCFT) when a Landau expansion of the free energy functional \cite{Leibler80} is performed and truncated after the quadratic term. This random phase approximation (RPA) is accurate in the linear response regime where density modulations are shallow. The dynamical (model B) equation for the relaxation of the Fourier components of the segment density reads \cite{Semenov86,qi2017dynamic, wang2019collective,muller2022memory}
\begin{equation}
    \frac{d\rho_{\bf k}(t)}{dt}=-\frac{{\bf k}^2\Lambda(k)}{S_{\bf k}(0)/N}\rho_{\bf k}(t).
\end{equation}
Agreement with the tICA generated dynamics is achieved when the Onsager coefficient $\Lambda(k)$ takes the usual wavevector dependent, time-local form for D-SCFT \cite{wang2019collective}, $\Lambda(k)=DS_{\bf k}(0)/N$.

\subsection{Hidden variables method for memory effects}

For  $t<\tau_R$, the decay of the dynamic structure factor of a Rouse chain is no longer purely exponential and reflects the subdiffusive intrachain relaxation. The Markovian model eq.~\eqref{eq-ticamodel} is not able to describe this dynamics with memory.
We now extend the Markovian dynamics with an approach based on hidden variables. First we choose a base lag time $\tau$. We can compute a finite dimensional approximation of the Koopman operator $K_1, K_2, K_3, \dots K_\rev{m}$ for each integer multiple of the lag time $\tau, 2\tau, 3\tau, \dots \rev{m}\tau$. Let's suppose our space of features is $N$ dimensional, so that all these matrices have size $N\times N$. If we assume Markov dynamics then $K_j = K_1^j$, but this is not necessarily true for systems with memory effects.

One approach we could take is to introduce hidden state of dimension $M$, not visible in the $N$ dimensional feature space, that encodes the memory of the system.  We might then hope to approximate the dynamics of the system with a $(N+M)\times(N+M)$ matrix, $A$ that acts on both the visible features and inferred hidden state.

\begin{figure}[t]
    \centering
    \includegraphics[width=0.9\linewidth]{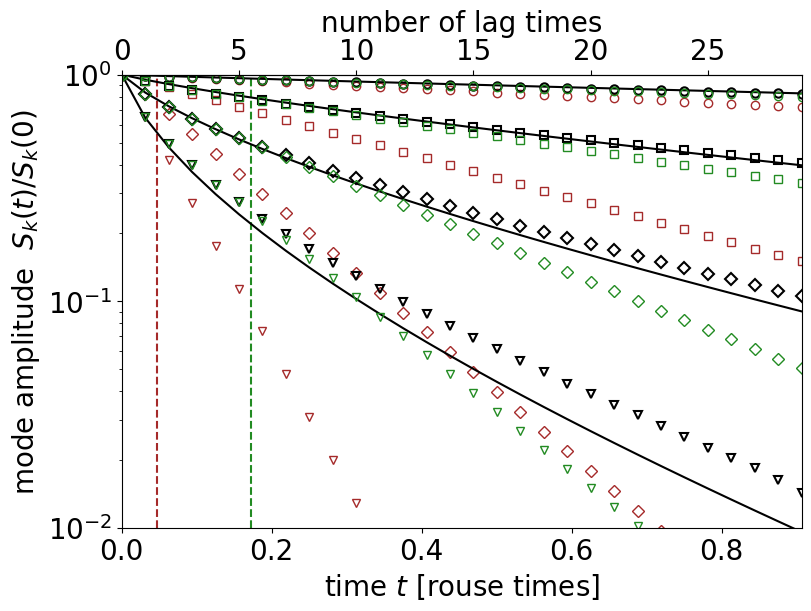}
    \caption{Single chain dynamical structure factor at wavevectors $kR_e = 2.38 (\circ), 4.76 (\square), 7.14 (\lozenge), 9.53 (\triangledown)$. Black symbols show the $S_k(t)/S_k(0)$ observed in the simulation while solid lines show the prediction of eq.~\eqref{dynsf-eq}. Brown symbols show the predictions of a simple Markovian Koopman model given by $\hat{\w{e}}_k^T K_1^{t/\tau}\hat{\w{e}}_k$ and green symbols are the predictions of the hidden Koopman model constructed from $K_1\dots K_5$ given by $\hat{\w{e}}_k^T P A^{t/\tau} P^T \hat{\w{e}}_k$, where $\hat{\w{e}}_k$ denotes the basis vector for spatial frequency $k$. The hidden Koopman model (green) depends on empirical time-lagged correlations to the left of the vertical green dashed line, but is not provided with time-lagged correlations to the right of it, and similarly for the simple Markovian model and the brown dashed line.}
    \label{fig:hiddenvar-test}
\end{figure}

\begin{figure}[t]
    \centering
    \includegraphics[width=0.9\linewidth]{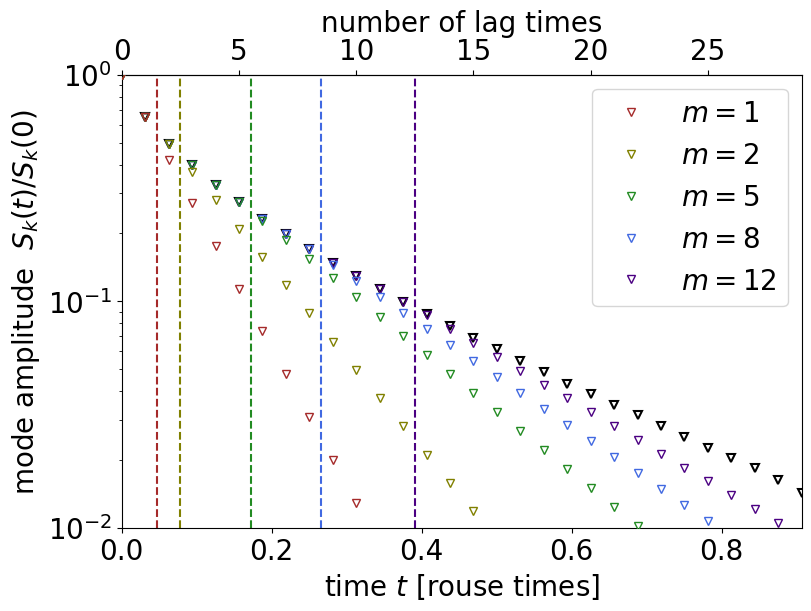}
    \caption{Single chain dynamical structure factor at wavevector $kR_e = 9.53$. Black symbols show the $S_k(t)/S_k(0)$ observed in the simulation while the rest show the predictions of a hidden Koopman model constructed from $K_1\dots K_\rev{m}$ (the legend shows the value of \rev{$m$}) given by $\hat{\w{e}}_k^T P A^{t/\tau} P^T \hat{\w{e}}_k$, where $\hat{\w{e}}_k$ denotes the basis vector for spatial frequency $k$. Vertical dashed lines separate lag-times used to construct $A$ from those where we are extrapolating. Note that $\rev{m}=1$ is equivalent to the simple Markovian model.}
    \label{fig:hiddenvar-test2}
\end{figure}

How might we find such a matrix? From observations of the system dynamics, it is possible to use tICA to determine $K_1, K_2, K_3, \dots K_\rev{m}$. One fairly simple option is to then let $M = (\rev{m}-1)N$ and set (see Appendix):

\begin{equation} \label{hiddenmat}
    A = \begin{pmatrix}
        K_1 && K_2 - K_1^2 &  \cdots \\
        I && 0 \\
        0 && I \\
        & \vdots && \ddots
    \end{pmatrix}
\end{equation}

This matrix directly remembers $n$ previous states of the density field, and the top row is constructed to ensure that $PA^jP^T = K_j$ for any $j = 1 \dots \rev{m}$, where $P$ projects to the $N$ visible features. To be explicit, $P$ is the following block matrix:

\begin{equation}
    P = \begin{pmatrix}
        I & 0 & 0 & \cdots \\
    \end{pmatrix}
\end{equation}

\rev{For times up to $m\tau$, the matrix $A$ is merely a repackaging into one object of the $m$ tICA analyses that were performed in order to construct it. However, for times greater than $m\tau$, it does generate predictions for lagged correlations that are genuinely new, in the sense that those correlations were not used in the construction of $A$.}

In order to test our proposed generalized Koopman model with hidden memory, we perform standard molecular dynamics simulations of an equilibrated bead-spring polymer melt. Our system contains 1,000 bead-spring chains of $N=20$ monomers each at the standard reduced density $\rho^*=0.85$ simulated at the constant melt temperature $T=1$ \cite{kremer1990dynamics}. These unentangled chains can be expected to be in the Rouse regime, but are subject to excluded volume effects. 

In Fig.~\ref{fig:hiddenvar-test}, we show the decay of the single chain dynamical structure factor for several wavevectors. Black symbols correspond to measurement in the simulation and are in good agreement with the  solid lines that show the ideal Rouse model prediction of eq.~\eqref{dynsf-eq}. The nonexponential decay at higher wavevectors is clearly visible. Green symbols show the predictions of our hidden variables model that has memory up to the green vertical dashed line. By construction, the model fully reproduces the memory effects up to that time. It appears to predict purely Markovian behavior (exponential relaxation) beyond that timescale. Brown symbols show instead the behavior of Koopman model without memory, which can only describe purely exponential relaxation for times greater than the lagtime $\tau$. The Rouse time for the system was estimated by computing the center-of-mass diffusion $D$ and the root-mean-square end-to-end distance $R_e$ from simulation data and using these to estimate $\tau_R$.

Figure ~\ref{fig:hiddenvar-test2} explores the effect of varying \pb{$m$} (the number of tICA analyses we perform to construct the hidden-variables model) for a selected wavevector. As \pb{$m$} increases, more and more of the nonexponential decay affected by memory is captured until the crossover to exponential relaxation is reached.

\subsection{Time-local formulation}
An alternative approach to treat memory effects in a time-local fashion consists in introducing time-dependent relaxation rates $R_{\bf kk^\prime}(t)$ via the rate of change of the equilibrium time-lagged covariances, ~\cite{dominic2023building} 
\begin{equation}
    \dot{C}_{t0}=R_{\bf kk^\prime}(t)C_{t0}
    \label{timelocal-eq}
\end{equation}
so that we can determine the rates from their derivatives:
\begin{equation}
    R_{\bf kk^\prime}(t)=\frac{\dot{C}_{t0}}{C_{t0}}.
\end{equation}
For the Rouse chain, we compute
\begin{eqnarray}
    R_{\bf kk}(t)&=&-\int_0^1 ds \int_0^1 ds^\prime  [{\bf k}^2D \nonumber\\ &-&\frac{2({\bf k}R_e)^2}{3\pi^2\tau_R}\sum_{p=1}e^{-p^2t/\tau_R}\cos(\pi ps)\cos(\pi p s^\prime)] \h\h.
\end{eqnarray}
As expected, $R_{\bf kk}(t\gg \tau_R)=-{\bf k}^2D$ is independent of time, signaling pure exponential decay in the long time limit. In general, the time-local method requires taking numerical derivatives, which can be prone to noise. An attractive alternative is therefore to instead formally integrate eq.~\eqref{timelocal-eq} to obtain
\begin{equation}
    C_{(t+\tau) 0}=K^T(t,t+\tau)C_{t0}
\end{equation}
where $K^T(t,\tau)=\exp[-\int_t^{t+\tau}dt^\prime R_{\bf kk^\prime}(t^\prime)]$ \cite{sayer2023compact,dominic2023building}. In practice, we compute the propagator $K^T$ through matrix inversion: 
\begin{equation}
K^T(t,\tau)=C_{(t+\tau)0}C_{t0}^{-1}.
\label{tdkoop-eq}
\end{equation}  It is now obvious that the propagator becomes a time-dependent tICA operator. The memoryless exponential decay of correlation functions recovers the bare (time-independent) tICA propagator eq.~\eqref{eq-ticamodel}.

In this formulation, the dynamics of the density modes is written in time-convolutionless form, 
\begin{equation}
    \frac{d\rho_{\bf k}(t)}{dt}=\rev{\sum_{k^\prime}}R_{\bf kk^\prime}(t)\rho_{\bf k^\prime}(t).
\end{equation}
We note that an equivalent trajectory can be generated by introducing a non-local memory kernel $\mathcal{M}_{kk^\prime}(t-t^\prime)$ and using the dynamical equation  \cite{sayer2023compact,dominic2023building}
\begin{equation}
    \frac{d\rho_{\bf k}(t)}{dt}=\dot{\rho}_{\bf k}(0)\rho_{\bf k}(t)+ \int_0^tdt^\prime \mathcal{M}_{kk^\prime}(t-t^\prime)\rho_{\bf k^\prime}(t^\prime)
\end{equation}
with $=\dot{\rho}_{\bf k}(0)$ as initial condition.   

\section{Phase separation of a diblock copolymer melt}
In equilibrium, all eigenmodes of the Koopman operator either decay to zero or become stationary. In many situations of interest, however, some modes grow with time before reaching an eventual steady state. A paradigmatic example is spinodal decomposition following a rapid temperature quench, where density fluctuations initially grow (not necessarily in a purely exponential fashion) and then saturate at late times. 

In a nonequilibrium situation, time-translation invariance no longer holds and the covariance matrices can no longer be computed by averaging along (equilibrium) trajectories. It appears natural to replace the time average with an average over an ensemble of (nonequilibrium) trajectories.

In order to capture nonlinear effects that mix normal modes, we consider a \rev{nonequilibrium} dynamical equation of the form 
\begin{equation}
     \frac{d\rho_{\bf k}^{(1)}(t)}{dt}=\sum_i\rev{\sum_{k^\prime}}R_{\bf kk^\prime}^{(i)}(t)\rho_{\bf k^\prime}^{(i)}(t)
\label{contmodel-eq}     
\end{equation}
where $\rho_{\bf k}^{(i)}(t)=1/V\int d^dr \exp(i{\bf k}{\bf r})\rho({\bf r},t)^i$ denotes the Fourier transform of powers \rev{$i$} of the segment density $\rho({\bf r})$. \rev{Eq.~\ref{contmodel-eq} may be viewed as an order parameter expansion of the density, were all powers permitted by symmetry can appear. The description can be expanded to include other hydrodynamic variables as needed.}

Multiplying with $\rho_{\bf -k^\prime}^{(j)}(0)$ and averaging yields (we omit the indices ${\bf kk^\prime}$ for the moment)
\begin{equation}
     \dot{C}_{t0}^{(1)(j)}=\sum_i R^{(i)}(t)C_{t0}^{(i)(j)}
    \label{timelocalgen-eq}
\end{equation}
\rev{
This equation can be formally integrated:
\begin{equation}
   C_{(t+\tau)0}^{(1)(j)}=e^{R^{(1)}\tau}C_{t0}^{(1)(j)}+e^{R^{(1)}(t+\tau)}\sum_{i>1}\int_t^{t+\tau}ds e^{-R^{(1)}s}R^{(i)}C_{s0}^{(i)(j)} 
\end{equation}
The integrals containing the nonlinear terms can be approximated and linearized for small lag time $\tau$ as follows:
\begin{equation}
\begin{aligned}
    &\sum_{i>1}\int_t^{t+\tau}ds e^{-R^{(1)}s}R^{(i)}C_{s0}^{(i)(j)} \approx\\
    &\sum_{i>1}\frac{1}{2}\left[e^{-R^{(1)}t}R^{(i)}\tau C_{t0}^{(i)(j)} + e^{-R^{(1)}(t+\tau)}R^{(i)}\tau C_{(t+\tau)0}^{(i)(j)} \right]=\\
    &\sum_{i>1}\frac{1}{2}\left[e^{-R^{(1)}t}R^{(i)}\tau C_{t0}^{(i)(j)}+ e^{-R^{(1)}(t+\tau)}R^{(i)}\tau C_{t0}^{(i)(j)}+\mathcal{O}(\tau^2) \right]
\end{aligned}
\end{equation}
}
After multiplying with the prefactor $e^{R^{(1)}(t+\tau)}$ and keeping only terms of order $\mathcal{O}(\tau)$, the rate coefficients can finally be obtained as the solution of a linear system of equations,
\begin{equation}
    \begin{pmatrix}
        C_{(t+\tau )0}^{(1)(1)} \\
        \vdots \\
        C_{(t+\tau )0}^{(1)(n)} \\
    \end{pmatrix} = \begin{pmatrix}
        C_{t0}^{(1)(1)}  & \cdots &  C_{t0}^{(n)(1)}   \\
        \vdots & \ddots & \vdots \\
         C_{t0}^{(1)(n)} & \cdots &  C_{t0}^{(n)(n)}
    \end{pmatrix}
    \begin{pmatrix}
        \exp[\tau R^{(1)}(t)] \\
        \vdots \\
        \tau R^{(n)}(t) \\
    \end{pmatrix}
\label{coeff-eq}    
\end{equation}
where $n$ denotes the highest power of the nonlinear terms. 

\begin{figure}
    \centering
    \includegraphics[width=1\linewidth]{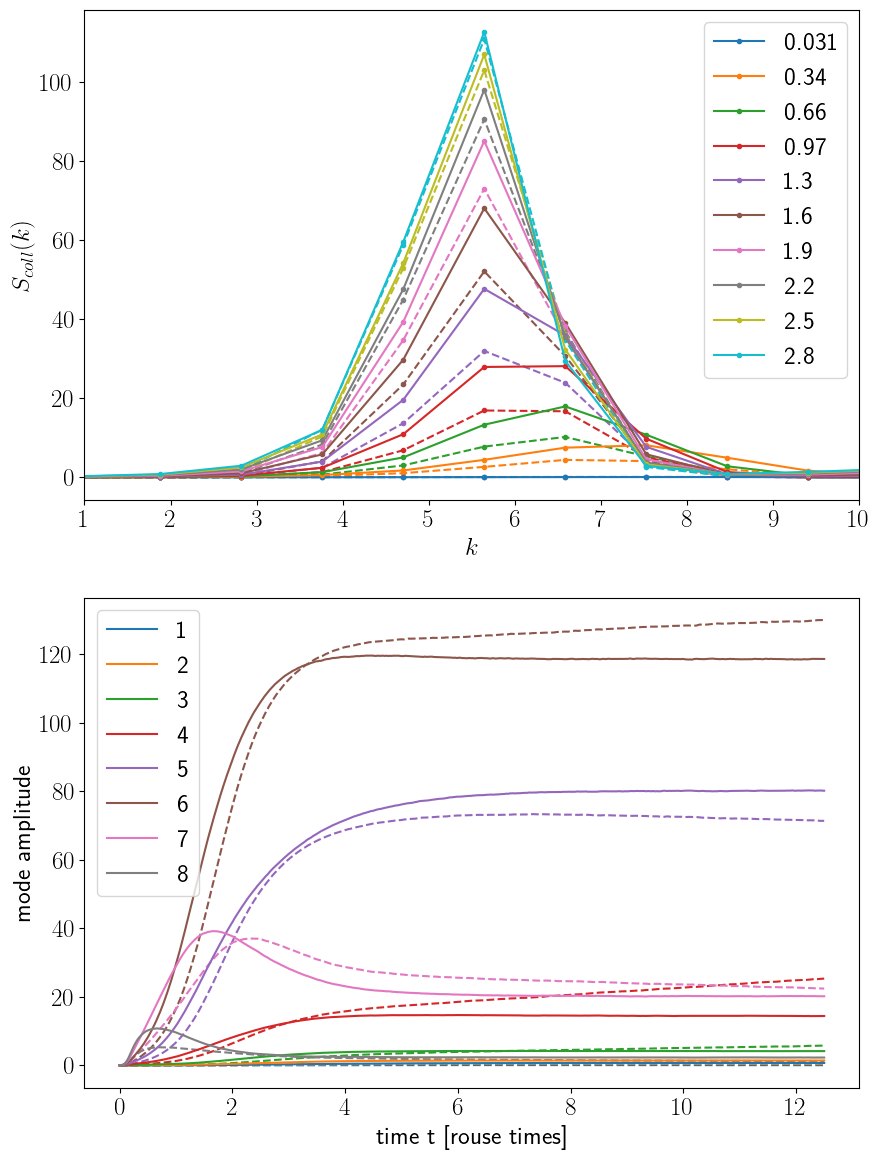}
    \caption{Top: Growth of the collective structure factor of a symmetric diblock copolymer system comprised of 4,000 chains undergoing phase separation after an instantaneous quench from $\chi N=0$ to $\chi N=40$ with $\kappa N=30$. Legend indicates elapsed Rouse time.  Bottom: Time evolution of low frequency modes. Solid lines SCMF simulations, dashed lines continuum model. Legend indicates mode index.}
    \label{fig:sq-dyn}
\end{figure}

In what follows, we consider the formation of ordered lamellae in a symmetric diblock copolymer melt \cite{matsen1994stable}. As microscopic reference system, we use particle-based Monte-Carlo simulations of a soft coarse-grained polymer model to study the short-time dynamics of polymer melts. The {SOMA} simulation package \cite{Schneider2019GPU} performs Monte-Carlo simulations with the Single-Chain-in-Mean-Field (SCMF) algorithm \cite{Daoulas2006Nov} in the canonical ensemble.

\rev{A total of $N_p$} macromolecules are described by a bead-spring model in three-dimensional space. A symmetric copolymer is comprised of $N=64$ coarse-grained segments; half of the segments, $fN=32$, are of type $A$ and the other half are of type $B$. The chain connectivity is represented by harmonic spring along the linear backbone of the flexible macromolecule
\begin{equation}
\frac{{\cal H}_{\rm b}}{k_{\rm B}T} = \sum_{i=1}^{N_p} \frac{3(N-1)}{2R_{e0}^2} \sum_{s=1}^{N-1}\left[  {\bf r}_i(s+1)-{\bf r}_i(s)\right]^2    
\end{equation}
where ${\bf r}_i(s)$ denotes the position of the $s^{{\rm th}}$ segment on the $i^{{\rm th}}$ copolymer. 

Non-bonded interactions are expressed via the particle-based densities
\begin{equation}
\hat \phi_A({\bf r}|\{{\bf r}_i(s)\}) = \frac{1}{\rho_0} \sum_{i=1}^{N_p}\sum_{s=1}^{fN} \delta({\bf r}-{\bf r}_i(s)) 
\label{eqn:dens}
\end{equation}
and likewise for the normalized $B$-segment, number density. The non-bonded interactions take the form
\begin{equation}
\begin{aligned}
\frac{{\cal H}_{\rm nb}}{k_{\rm B}T} = \frac{\rho_0}{N}\int\frac{{\rm d}{\bf r}}{R_{e0}^3} & \{ - \frac{\chi N}{4} \left[\hat \phi_A -\hat \phi_B\right]^2 \\ 
& +  \frac{\kappa N}{2} \left[ \hat \phi_A + \hat \phi_B-1\right]^2 \}    \label{eqn:Hnb}
\end{aligned}
\end{equation}

\begin{figure}[t]
    \centering
    \includegraphics[width=1\linewidth]{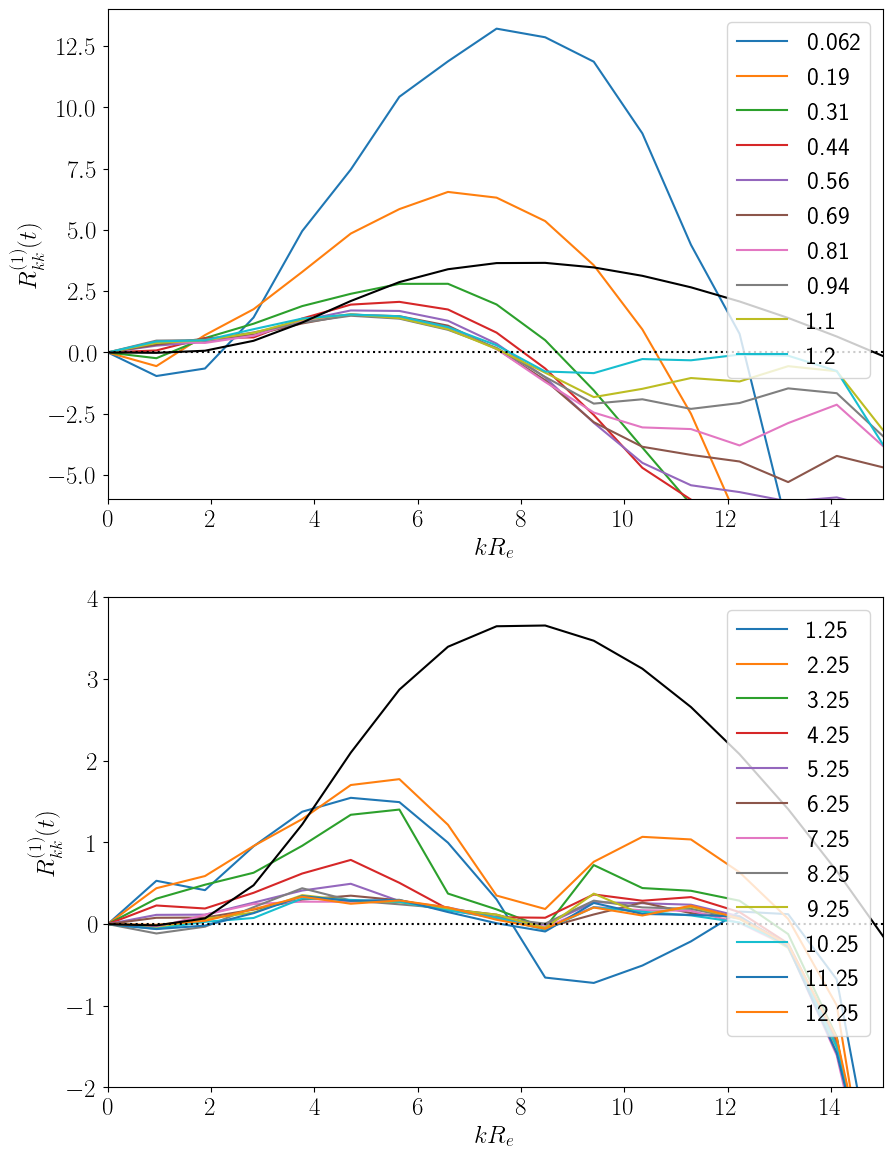}
    \caption{Linear rate coefficient at different simulation times (in units of $\tau_R$) indicated in the legend. Black line shows the rate $-k^2\Lambda(k)[N/S_k(0)-2\chi N]$ predicted by the quadratic term in an RPA expansion of the free energy.}
    \label{fig:r1}
\end{figure}

As proof of concept, we consider a quasi-onedimensional simulation box of dimensions $6.68\R \times 0.83\R \times 0.83\R $ and gather statistics over 1,000 trajectories. In our continuum model, we consider only odd terms since even terms vanish by symmetry and set $n=3$. In this simple case, the covariance matrices are diagonal and it is sufficient to consider only the wavevector in the long direction of the simulation box to probe the formation of periodic structures. Rate coefficients  $R^{(1)}_{kk}(t)$ and $R^{(3)}_{kk}(t)$ are obtained by solving eq.~\ref{coeff-eq} with a lag time of $1/32\, \tau_R$, and the continuum model eq.~\ref{contmodel-eq} is solved using a simple Euler method and starting from the same initial field configurations as in the SCMF simulations.

Figure \ref{fig:sq-dyn} compares the time evolution of the structure factor of the polymer system for the two models. In the \rev{chosen} geometry, lamella structures with 5 and 6 periods are the most common, followed by a few systems exhibiting 4 and 7 periods. It can be appreciated that the learned continuum model reproduces the dynamics quite well. At early times, spinodal decomposition is faster in the particle based model since it has the ability to overcome small free energy barriers during the quench. For this reason, exact agreement between the models cannot be expected. At late times, the continuum model predicts on average a higher periodicity than the SCMF simulations.

\begin{figure}[t]
    \centering
    \includegraphics[width=1\linewidth]{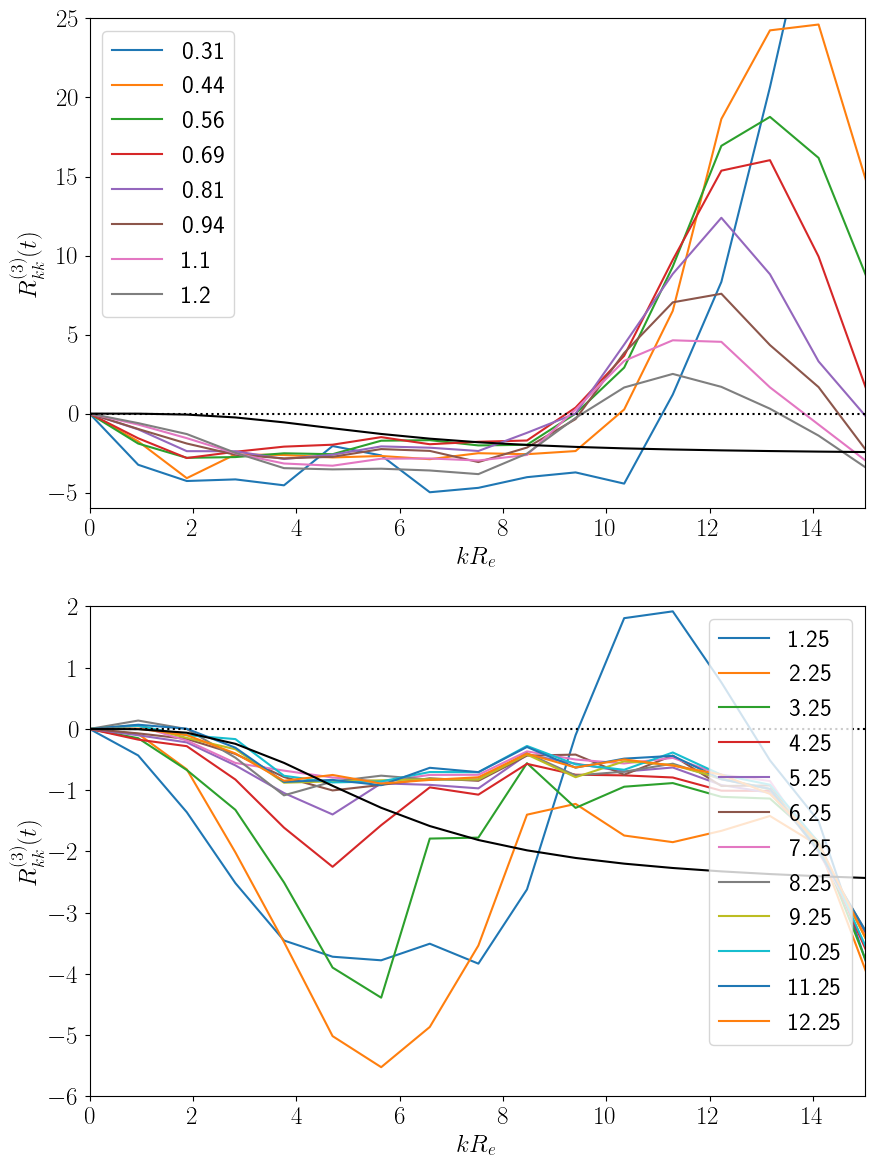}
    \caption{Cubic rate coefficient at different simulation times (in units of $\tau_R$) indicated in the legend. Black line shows the rate $-k^2\Lambda(k)[156.56/6]$ predicted by the quartic term in an RPA expansion of the free energy.}
    \label{fig:r3}
\end{figure}

\begin{figure}[t]
    \centering
    \includegraphics[width=1\linewidth]{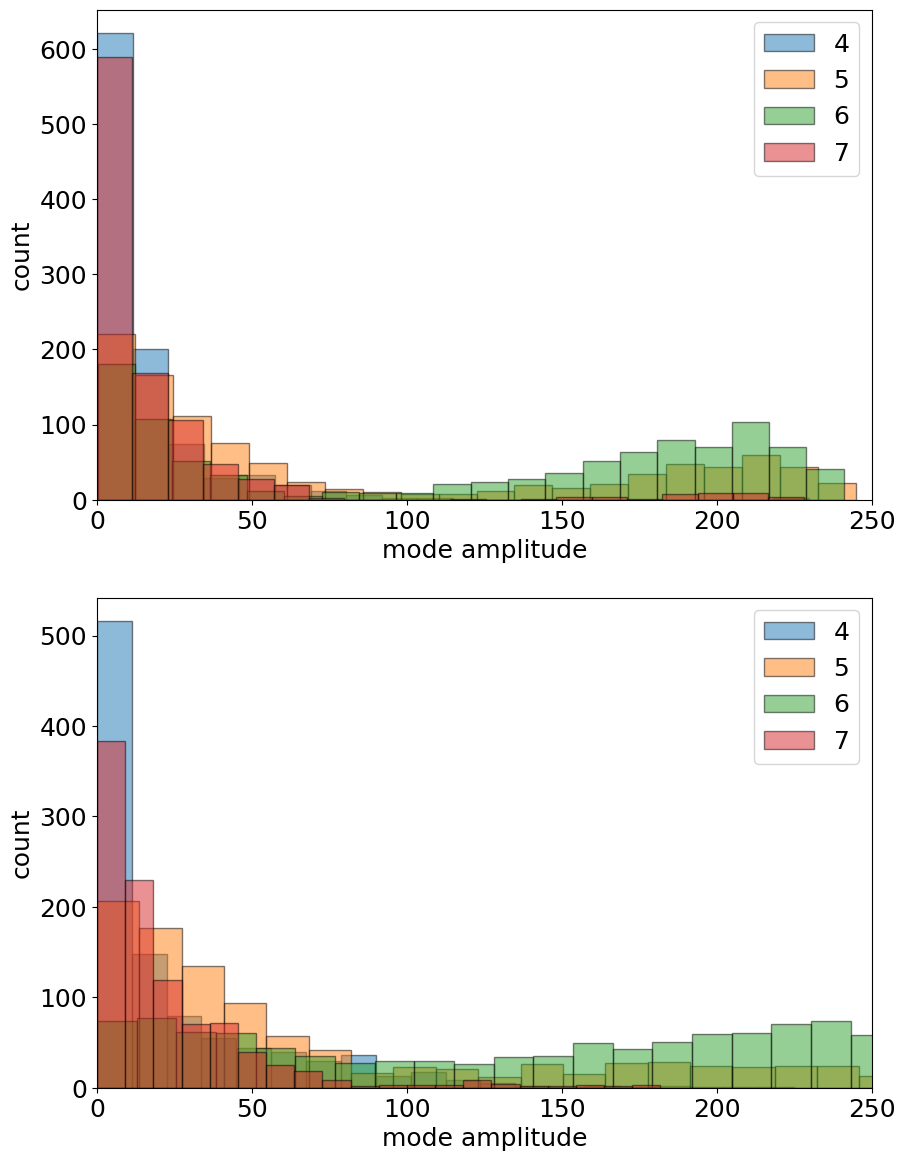}
    \caption{Distribution of final morphologies. Top panel SCMF simulations, bottom panel continuum model.}
    \label{fig:hist}
\end{figure}

Figure \ref{fig:r1} shows the wavevector dependence of the rate coefficent $R^{(1)}_{kk}(t)$ multiplying the linear term for different times elapsed since the quench. At all times, some modes have positive rates and their amplitude therefore grows. Three dynamical regimes can be identified: For short times less than approx $0.5\tau_R$,  the rate $R^{(1)}_{kk}(t)$ is large and positive and changes rapidly with time, reflecting memory effects \cite{steffen2025collective}. For $0.5 \tau_R \leq t < 2 \tau_R$, the rate does not change with time, which means that the modes exhibit simple exponential growth. Finally, for times larger than approx.  $5 \tau_R$, the rates again no longer change with time, reflecting the saturation of the growth dynamics. We also show for comparison the (time-independent) growth rates that would be used in a D-SCFT model with RPA approximation \cite{Leibler80,muller2018continuum}; these do not agree well with the simulations. 

The rate coefficient $R^{(3)}_{kk}(t)$ multiplying the nonlinear term is shown in Figure \ref{fig:r3}. In general, these rates tend to have the opposite sign of $R^{(1)}_{kk}(t)$. For instance, at early times $t \leq 1 \tau_R$ $R^{(3)}_{kk}(t)>0$ for larger $kR_e>10$, while $R^{(1)}_{kk}(t)<0$ in the same range of wavevectors. At late times $R^{(3)}_{kk}(t)<0$ for all values of $k$ so that the positive linear rates can be balanced and a steady state is achieved.  A convergence to the RPA expression \cite{Leibler80,muller2018continuum} can be observed for $kR_e\leq 5$ in that steady state.

How do the final morphologies generated by the continuum model compare to those of the reference SCMF simulations? This question is answered by Figure \ref{fig:hist}, where we compare the full distributions of mode amplitudes corresponding to 4, 5, 6 or 7 lamellae between particle based SCMF simulations (top panel) and continuum simulations (bottom panel). The agreement is reasonable keeping in mind that the continuum simulations evolve a distribution of random initial conditions deterministically. 

\section{Conclusions}
Data driven approaches for the analysis of the dynamics of \rev{(bio)macromolecules} are attracting interest due to their ability to approach the system with minimal a priori assumptions. For polymers undergoing Rouse dynamics, we first recalled that the bases of largest variance given by PCA and slowest modes given by Rouse analysis coincide. The application of tICA to single chain and collective density fluctuations prescribes a memoryless relaxation dynamics of the modes, equivalent to "model B" dynamics for conserved fields. We then introduced a possibility to introduce memory effects in the tICA formalism via hidden variables and constructed an effective continuum model for spinodal decomposition from microscopic reference simulations that accounts for these memory effect. Our time-local formulation avoids the use of memory kernels and convolution integrals, but mixes thermodynamics and kinetics in time-dependent rate coefficients. Although this requires separate reference simulations for each set of thermodynamic parameters, we expect that this formalism can be fruitfully combined with ongoing efforts on dynamic coarse graining of polymer systems \cite{rottler2020kinetic,mantha2020bottom,schmid2020dynamic,Li_2021} and extend them to complex chain topologies.

\section*{Acknowledgements}
We thank Marcus M\"uller for helpful comments on the manuscript. Financial support from the discovery grant program of the Natural Sciences and Engineering Research Council of Canada (NSERC) is gratefully acknowledged.

\section*{Data availability statement}
Source code for the hidden variables method can be found at \pb{\href{https://github.com/pb1729/hidden-koopman}{\texttt{https://github.com/pb1729/hidden-koopman}}}.

\section*{Appendix: Computing the top row of the block matrix in equation \ref{hiddenmat}}

We would like to construct a matrix $A$ such that for all $1\leq j \leq n$,

\begin{equation}\label{upperleftconstr}
    P A^j P^T = K_j
\end{equation}

As an ansatz, we take

\begin{equation}
    A = \begin{pmatrix}
        R_1 & R_2 & R_3 & R_4 & \cdots \\
        I  \\
        & I \\
        & & I \\
        & & & \ddots
    \end{pmatrix}
\end{equation}

where the top row is left unspecified for the moment. If we take powers of this matrix, the following relationship between the $R_i$ and the $K_i$ must hold: For all  $1\leq j \leq n$,

\begin{equation}
    PA^j P^T = R_j + \sum_{k=1}^{j-1} R_{j-k} PA^{k}P^T
\end{equation}

and if we require $A$ to satisfy equation \ref{upperleftconstr}, then:

\begin{equation} \label{RKrelate}
    K_j = R_j + \sum_{k=1}^{j-1} R_{j-k} K_k
\end{equation}

We can now create generating functions i.e. write the $R_i$ and the $K_i$ as individual elements of $\left(\CC[z]/(z^{n+1})\right)^{N\times N}$. (Recall that $N$ was the size of the individual blocks in our block matrix.)

\begin{eqnarray}
    K(z) = \sum_{j=1}^n K_j z^j \\
    R(z) = \sum_{j=1}^n R_j z^j
\end{eqnarray}

In this form, equation \ref{RKrelate} becomes:

\begin{equation}
    K(z) = R(z) + R(z)K(z)
\end{equation}

We can then solve for $R(z)$:

\begin{equation}
    R(z) = K(z)(I + K(z))^{-1}
\end{equation}

For ease of computing $R(z)$, these generating functions can be expanded back out into block-Topelitz matrices. For example, with $n=3$, we'd have:

\begin{equation}
\text{
\scalebox{0.8}{
    ${\begin{pmatrix}
        0\\
        R_1 & 0 \\
        R_2 & R_1 & 0\\
        R_3 & R_2 & R_1 & 0
    \end{pmatrix}
    = \begin{pmatrix}
        0\\
        K_1 & 0 \\
        K_2 & K_1 & 0\\
        K_3 & K_2 & K_1 & 0
    \end{pmatrix}\begin{pmatrix}
        I\\
        K_1 & I \\
        K_2 & K_1 & I\\
        K_3 & K_2 & K_1 & I
    \end{pmatrix}^{-1}}$
}}
\end{equation}

These expressions can then be evaluated by a computer algebra package, leaving the $K_i$'s as free symbols. Because these matrices are lower triangular, as we increase $n$ and add more $K_i$'s to the end of our list, the list of $R_i$'s likewise gets longer, but existing entries do not change. So here are the first few $R_i$, as determined by computer:

\begin{equation}
    \begin{matrix}
        R_1 \h = & K_1 \\
        R_2 \h = & K_2 - K_1^2 \\
        R_3 \h = & K_3 - K_2K_1 - K_1K_2 + K_1^3 \\
        R_4 \h = & K_4 - K_3K_1 - K_2^2 + K_2K_1^2 - K_1K_3 \\
            & + K_1K_2K_1 + K_1^2K_2 - K_1^4 \\
            \vdots
    \end{matrix}
\end{equation}

If we look for patterns, we can see that there is one term in $R_j$ corresponding to each ordered partition of $j$ (there are $2^{j-1}$ of these), and the coefficient is either $-1$ or $+1$ depending on whether the partition has an even or odd number of parts.\\

\end{document}